\documentclass[prl,twocolumn,aps,superscriptaddress]{revtex4-2}

\usepackage{amsmath,amssymb,mathtools}
\usepackage{bm}
\usepackage{booktabs}
\usepackage{graphicx}
\usepackage{hyperref}
\usepackage{xcolor}

\newcommand{\Trev}{\hat\Theta}
\newcommand{\bs}{\boldsymbol}

\newcommand{\Ham}{\hat{H}}

\providecommand{\ket}[1]{|#1\rangle}

\begin{document}

\title{Spin Rotatory Strength as the Equilibrium Observable\\
for Chiral-Induced Spin Selectivity}

\author{Rub\'en D. Guerrero}
\email{rudaguerman@gmail.com}
\affiliation{NeuroTechNet S.A.S., 1108831, Bogot\'a, Colombia}
\affiliation{Quantum and Computational Chemistry Group (QCCG),
Universidad Nacional de Colombia, Bogot\'a, Colombia}

\date{\today}

\begin{abstract}
We identify the spin rotatory strength---a chirality-odd,
finite-frequency spin--dipole cross response---as the
equilibrium observable for chiral-induced spin selectivity.
Exact diagonalization of a Kane--Mele--Hubbard model
($N{=}4,6,8$) reveals that gapped molecules robustly
suppress the response, whereas near-degenerate systems show
amplification at small $N$. Three size-independent results
are exact: the static response vanishes by time-reversal
symmetry, nonzero SU(2) Wilson-loop flux is a necessary
condition, and the response has the structure of natural
optical activity.
\end{abstract}

\maketitle


Electrons traversing chiral molecules exhibit spin polarizations far
larger than spin-orbit coupling (SOC) alone can
explain---a phenomenon termed chirality-induced spin selectivity
(CISS)~\cite{NaamanWaldeck2012,NaamanWaldeck2015}. Room-temperature spin
polarizations exceeding 60\% have been measured in helicene monolayers, DNA
self-assembled films, and oligopeptide
junctions~\cite{Gohler2011,Xie2011,Kettner2015,Mishra2013}, yet the atomic SOC of
carbon, nitrogen, and oxygen is only $\sim$10--30~meV---orders of magnitude too
small for a first-order SOC perturbative estimate.

Prior theoretical work has taken two broad routes. Transport-based
approaches~\cite{GuoSun2012,Medina2012,Matityahu2016,Michaeli2016} model CISS
through spin-dependent transmission in open circuits with leads and dephasing,
producing nonzero spin polarization but requiring a
nonequilibrium steady state. First-principles calculations with SOC
systematically undershoot experimental spin polarizations by orders of
magnitude~\cite{DalumHedegard2019}, while vibronic coupling has been proposed as
an amplifier~\cite{Fransson2019,Fransson2020}. As Naaman and
Paltiel~\cite{NaamanPaltiel2026} have stressed, single-electron models are
qualitatively insufficient and the observed correlation between spin
polarization and chiroptical response points to missing many-body
ingredients. No framework sharply distinguishes when equilibrium mechanisms
suffice from when transport is required.

Here we establish three results. (i)~The equilibrium
spin-magnetoelectric response has the structure of natural optical activity
(NOA): a chirality-odd spin rotatory strength that is nonzero at finite
frequency without dissipation, while the static response vanishes by
time-reversal symmetry. (ii)~A nonzero SU(2) Wilson-loop flux on the molecular
backbone is an independent necessary condition, ruling out all tree-connected
(open-chain) models. (iii)~Exact diagonalization reveals a two-regime
structure: gapped molecules robustly suppress the spin rotatory strength,
while near-degenerate (diradicaloid, i.e., near-degenerate
singlet--triplet) systems show amplification that is strong at small
system sizes.

\section{The spin rotatory strength}

\textit{Static no-go.}---Consider a time-reversal-invariant molecule
($\Trev\Ham\Trev^{-1} = \Ham$) with ground state $\ket{0}$. The static
spin-magnetoelectric polarizability $\alpha_{kz} =
d\langle\hat{S}_k\rangle / dE_z |_0$ couples the $\Trev$-odd spin
operator to the $\Trev$-even electric field. For even-electron systems
(the typical molecular case), Kramers' theorem does not apply and the
ground state is generically nondegenerate (accidental degeneracies
are possible but nongeneric and immediately lifted by any
perturbation). Time reversal then gives
$\Trev\ket{0} = e^{i\phi}\ket{0}$, which directly implies
$\langle 0|\hat{S}_k|0\rangle = 0$ and, through the sum-over-states
expression,
\begin{equation}
\alpha_{kz}(0) = 0, \qquad \forall\,\zeta_{\mathrm{SOC}}.
\label{eq:nogo}
\end{equation}
For odd-electron systems with Kramers-degenerate ground states
$\{\ket{0},\Trev\ket{0}\}$, the antiunitary identity
$\langle\Trev\psi|\hat{O}|\Trev\phi\rangle = \varepsilon_O
\langle\psi|\hat{O}|\phi\rangle^*$ (with $\varepsilon_{S_k} = -1$,
$\varepsilon_z = +1$) flips the sign of each spectral contribution while
preserving the energies; summing the linear-response contributions
from both Kramers partners then gives the same result. The static response therefore vanishes for both
even- and odd-electron systems---an exact symmetry theorem, not a
perturbative statement~\cite{LandauLifshitz}.

\textit{Finite-frequency response.}---At finite frequency, the
retarded Kubo cross response
\begin{equation}
\chi_{S_k,z}(\omega) = \sum_{n\neq0} \left[
\frac{M_n}{\omega - \omega_n + i0^+} -
\frac{M_n^*}{\omega + \omega_n + i0^+}
\right],
\label{eq:chi}
\end{equation}
with $M_n = \langle 0|\hat{S}_k|n\rangle\langle n|\hat{z}|0\rangle$, is
generically nonzero whenever SOC and chirality are both present. Because $\hat{S}_k$ is $\Trev$-odd and $\hat{z}$ is
$\Trev$-even, $\Trev$-invariance of $\ket{0}$ and $\ket{n}$
forces $M_n^* = -M_n$, so $M_n$ is purely imaginary. Writing
$M_n = iR_n^{\mathrm{spin}}$ with
$R_n^{\mathrm{spin}} \in \mathbb{R}$ defines the \textit{spin rotatory
strength}---the residue of each spectral pole:
\begin{equation}
R_n^{\mathrm{spin}} = \mathrm{Im}\!\left[
\langle 0|\hat{S}_k|n\rangle\langle n|\hat{z}|0\rangle
\right].
\label{eq:Rspin}
\end{equation}
The absorptive spectral response
$\sigma_{kz}(\omega) = \pi \sum_n R_n^{\mathrm{spin}}
[\delta(\omega - \omega_n) + \delta(\omega + \omega_n)]$
is nonzero at any finite $\omega$ and vanishes only in the strict DC
limit, recovering Eq.~\eqref{eq:nogo}.

This quantity is the exact spin-sector analog of the optical rotatory
strength
$R_n = \mathrm{Im}[\langle 0|\hat{\bs\mu}_e|n\rangle \cdot \langle
n|\hat{\bs{m}}|0\rangle]$ in NOA~\cite{Barron2004}: in both cases, an
opposite-$\Trev$-parity pair of operators yields a purely imaginary transition
moment, and the absorptive response is $\mathrm{Re}\,\chi$~\cite{Rikken1997}.
Electron-vibron coupling broadens the $\delta$-functions into
Lorentzians~\cite{Fransson2019,Fransson2020} but is not required:
$R_n^{\mathrm{spin}}$ is finite at vanishing linewidth.

Two experiments can probe the spin rotatory strength directly:
magnetochiral dichroism~\cite{Rikken1997}
and spin-polarized photoabsorption spectroscopy, where
the cross-section difference for opposite spin orientations is
proportional to $\sigma_{kz}(\omega)$.

\section{SU(2) gauge structure}

An independent necessary condition constrains which molecular
topologies can support nonzero $R_n^{\mathrm{spin}}$. We model SOC as
SU(2)-valued bond hopping operators $U_{ij} \in \mathrm{SU}(2)$ on the
molecular backbone (the Kane--Mele
construction~\cite{KaneMele2005}). If the backbone graph is simply
connected (a tree), site-local SU(2) gauge rotations
$V_i$ exist such that $V_i U_{ij} V_j^\dagger = \mathbf{1}$ on every bond.
The transformed Hamiltonian then commutes with total spin, so all spin
matrix elements vanish:
\begin{equation}
R_n^{\mathrm{spin}} = 0, \qquad
\sigma_{kz}(\omega) \equiv 0, \qquad \forall\,\zeta_{\mathrm{SOC}}.
\label{eq:fluxnogo}
\end{equation}
A nonzero equilibrium response requires a multiply connected backbone
enclosing nonzero gauge-invariant SU(2) Wilson-loop flux
$\Phi_{\mathrm{SU(2)}} = \arccos\!\bigl(\tfrac{1}{2}\,\mathrm{tr}\,\mathcal{P}
\prod_\square U_{ij}\bigr) \neq 0$ (where $\mathcal{P}$ denotes
path-ordering)~\cite{Wilson1974}.

This result constrains the class of theoretical models, not molecular
candidates: most organic chromophores contain cyclic $\pi$-systems and
satisfy $\Phi_{\mathrm{SU(2)}} \neq 0$ for any nonzero SOC. The practical
consequence is that the widely studied nearest-neighbor open-chain SOC
models~\cite{GuoSun2012} sit at $\Phi_{\mathrm{SU(2)}} = 0$ and produce
nonzero CISS only through leads, dephasing, or real magnetic
flux---all requiring $\Trev$-breaking. The Wilson-loop criterion thus
provides an independent reason, beyond the static no-go, why such
models give identically zero in equilibrium.

\section{Exact diagonalization}

To verify these constraints and probe the regime structure, we perform
exact diagonalization of a Kane--Mele--Hubbard ring with $N$ sites
at half filling ($N$ electrons in $2N$ spin orbitals),
\begin{align}
\Ham &= -\tau \sum_{\langle ij\rangle}
c_{i\varsigma}^\dagger \left[e^{i(\gamma/2)\hat{n}_{ij}\cdot\bs\sigma}
\right]_{\varsigma\varsigma'} c_{j\varsigma'}
- \tau_2 \sum_{\langle\langle ij\rangle\rangle}
c_{i\varsigma}^\dagger c_{j\varsigma}
+ \mathrm{H.c.} \notag \\
&\quad + U_\mathrm{H} \sum_j n_{j\uparrow} n_{j\downarrow}
+ \sum_j \varepsilon_j n_j,
\label{eq:model}
\end{align}
where $\tau$ ($\tau_2$) is the (next-)nearest-neighbor hopping,
$\gamma$ is the SOC mixing angle with bond-chirality axis
$\hat{n}_{ij}$ (whose azimuthal increment $\Delta\varphi$ around
the ring encodes the helix pitch), $U_\mathrm{H}$ is the Hubbard
repulsion, and $\varepsilon_j$ are inversion-breaking on-site
potentials; the ring geometry ensures nonzero
$\Phi_{\mathrm{SU(2)}}$ while preserving $\Trev$ invariance. We study $N=4$, 6, 8 (dimensions 70, 924, 12\,870);
reference parameters: $\tau{=}1$, $\tau_2{=}0.4$, $\gamma{=}0.6$,
$\Delta\varphi{=}0.7$, $U_\mathrm{H}{=}1$
($\Phi_{\mathrm{SU(2)}}{=}0.677$). The 875-point $N{=}4$ scan
varies $\gamma{\in}[0.1,1]$, $\tau_2{\in}[0.1,0.8]$,
$U_\mathrm{H}{\in}[0.5,3]$, $\Delta\varphi{\in}[0.3,1]$.
Full diagonalization is used for $N{\leq}6$, ARPACK Lanczos for
$N{=}8$, with cross-validation at $\delta \leq 10^{-14}$.

\begin{figure*}[t]
\includegraphics[width=0.95\textwidth]{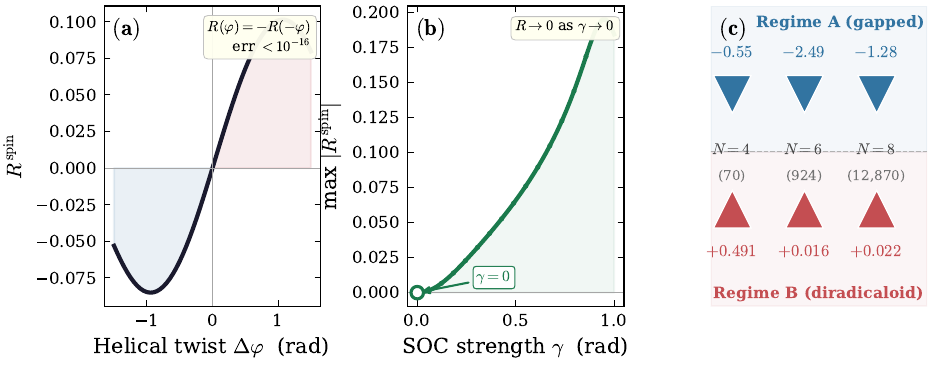}
\caption{Symmetry constraints and finite-size scaling of the spin
rotatory strength $R^{\mathrm{spin}}$.
(a)~Chirality-odd antisymmetry: $R(\varphi)=-R(-\varphi)$ holds to
machine precision (error $<10^{-16}$), confirming that the response
changes sign between enantiomers.
(b)~SOC dependence: $R^{\mathrm{spin}} \to 0$ as the SOC strength
$\gamma \to 0$, demonstrating that SOC is necessary but not
sufficient---a nonzero Wilson-loop flux is additionally required.
(c)~Power-law exponent $q$ (defined by
$R^{\mathrm{spin}} \propto \Delta^{-q}$ or
$\delta_{\mathrm{ST}}^{-q}$) across system sizes $N{=}4,6,8$:
Regime~A (gapped, blue) gives $q_A<0$ (suppressed) at all sizes;
Regime~B (diradicaloid, red) gives $q_B>0$ (amplified) but the
magnitude decreases from $+0.49$ ($N{=}4$) to $+0.02$ ($N{=}6,8$),
with the latter not statistically distinguishable from zero
(Table~\ref{tab:regimes}).
\label{fig:symmetry}}
\end{figure*}

\textit{Symmetry verification.}---All constraints hold at machine
precision across the full parameter scan (875 parameter sets for $N=4$):
$\mathrm{Re}\,\chi(0) < 3 \times 10^{-16}$ (static no-go);
$R^{\mathrm{spin}} < 7 \times 10^{-32}$ when SOC is removed;
$R^{\mathrm{spin}} < 2 \times 10^{-17}$ when chirality is removed;
identically zero on a tree backbone; and exact chirality-odd
antisymmetry $R(\varphi) = -R(-\varphi)$ [Fig.~\ref{fig:symmetry}(a),
error $< 10^{-16}$].  The response vanishes continuously as the
SOC strength $\gamma \to 0$ [Fig.~\ref{fig:symmetry}(b)], confirming
that SOC is necessary but not sufficient---the Wilson-loop topology
is the additional requirement.

\begin{figure*}[t]
\includegraphics[width=0.95\textwidth]{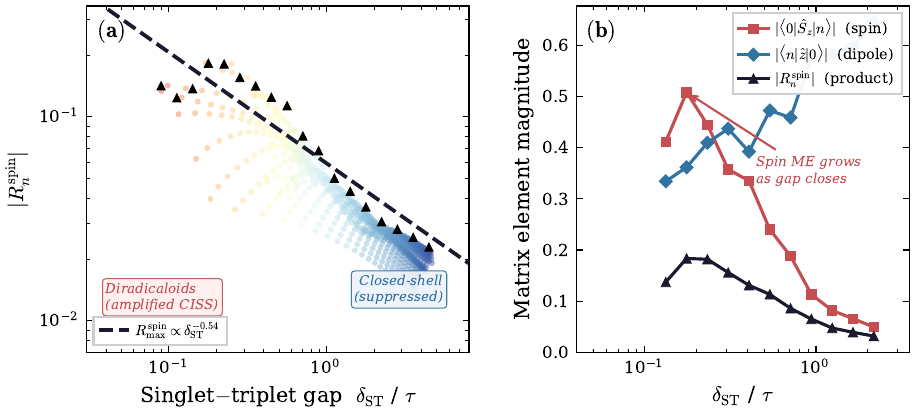}
\caption{Spin rotatory strength $|R^{\mathrm{spin}}_n|$ in the
diradicaloid regime (Regime~B; $N{=}4$ sites, 875 parameter sets).
(a)~$|R^{\mathrm{spin}}_n|$ vs.\ singlet--triplet gap
$\delta_{\mathrm{ST}}$; each point represents one parameter set, with
color encoding gap magnitude. The upper envelope follows a power law
$\propto \delta_{\mathrm{ST}}^{-0.54}$ (dashed line); the envelope exponent ($0.54$) is steeper than the all-points
regression ($q_B = 0.49$, Table~\ref{tab:regimes}) because the
envelope selects parameter sets with optimal dipole alignment. (b)~Mechanism decomposition along the
envelope: the spin matrix element
$|\langle 0|\hat{S}_z|n\rangle|$ grows as $\delta_{\mathrm{ST}}$
decreases while the dipole factor $|\langle n|\hat{z}|0\rangle|$ remains
bounded, confirming the SOC admixture mechanism of
Eq.~\eqref{eq:admixture}. This decomposition holds at $N{=}4$; at
$N{=}8$ the dipole factor becomes the dominant contributor (see text).
\label{fig:prediction}}
\end{figure*}

\textit{Two-regime structure.}---Systematic scans over independent
kinematic knobs at fixed $\Phi_{\mathrm{SU(2)}}$ reveal two regimes
(Table~\ref{tab:regimes}, Fig.~\ref{fig:prediction}).

\textit{Regime~A (gapped).}---We vary the orbital gap (the
single-particle HOMO--LUMO splitting) through spin-independent on-site
potentials $\varepsilon_j$ (linear, staggered, and cosine patterns with
amplitude $0$--$4\tau$) at fixed SOC and $U_\mathrm{H}$.  The spin rotatory
strength decreases as the gap closes ($q_A < 0$) for all three system
sizes (Table~\ref{tab:regimes}). Since only $\varepsilon_j$ varies in these scans at
fixed SOC, the suppression reflects charge-sector physics: the
gap-closing rearrangements reduce the peak spin--dipole spectral weight
faster than the increased SOC admixture ($\propto\zeta/\Delta$) can
compensate. Gapped-molecule CISS is therefore
weak in equilibrium, and large experimental signals in such systems
must arise from
transport~\cite{GuoSun2012,Medina2012,Matityahu2016,Fransson2020},
consistent with the DFT-SOC undershoot~\cite{DalumHedegard2019}.
The robust conclusion is $q_A < 0$ at all sizes; the specific exponent
is non-monotone ($-0.55$, $-2.49$, $-1.28$ for $N=4,6,8$), reflecting
sensitivity to the many-body spectrum that does not affect the sign.

\textit{Regime~B (near-degenerate).}---We drive the singlet--triplet
gap $\delta_{ST}$ toward zero by tuning the scalar next-nearest-neighbor
hopping $\tau_2$ at fixed $\Phi_{\mathrm{SU(2)}}$.
At $N=4$ (dim~$=70$), the spin rotatory strength grows as $\delta_{ST}$
decreases, with exponent $q_B = +0.49 \pm 0.02$ (95\% CI; $r=-0.99$)
[Fig.~\ref{fig:prediction}(a)]. The mechanism decomposition
[Fig.~\ref{fig:prediction}(b)] reveals what drives this amplification:
the spin matrix element $|\langle 0|\hat{S}_z|n\rangle|$ grows near
the singlet--triplet degeneracy through SOC admixture,
\begin{equation}
\langle 0|\hat{S}_z|n\rangle \propto
\frac{\langle T_1|\hat{H}_\mathrm{SOC}|S_0\rangle}{\delta_{ST}},
\label{eq:admixture}
\end{equation}
while the dipole factor $|\langle n|\hat{z}|0\rangle|$ remains bounded.
Near exact degeneracy, two-state hybridization with mixing angle
$\tan 2\vartheta = 2\langle T_1|\hat{H}_\mathrm{SOC}|S_0\rangle /
\delta_{ST}$ causes sub-linear saturation, explaining why $q_B < 1$:
at exact degeneracy the admixture saturates at $\sin 2\vartheta = 1$,
capping the spin matrix element.

At larger sizes, the amplification weakens substantially:
$q_B = +0.02 \pm 0.12$ ($N=6$) and $q_B = +0.02 \pm 0.15$ ($N=8$).
These confidence intervals include zero, so the amplification is
not statistically significant at $N = 6$ or $8$
[Fig.~\ref{fig:symmetry}(c)]. The mechanism decomposition also
shifts: at $N=8$, the dipole factor rather than the spin factor drives
the response. The two-state admixture of Eq.~\eqref{eq:admixture}
assumes a single dominant $T_1$ state; at $N\geq6$, the growing
triplet-state density dilutes the per-state SOC admixture, shifting
the balance to the dipole factor.

At $N=6,8$, $q_B$ is indistinguishable from zero within 95\% CI;
whether any amplification survives in the thermodynamic limit remains
open and requires larger-scale calculations or complementary analytical
approaches.

\begin{table}[b]
\caption{Finite-size scaling of the two-regime structure for
$N$-site Kane--Mele--Hubbard rings (Hilbert-space dimension ``dim'').
The exponent $q$ is defined by
$R^{\mathrm{spin}} \propto \Delta^{-q}$ (Regime~A, orbital gap) or
$\propto \delta_{ST}^{-q}$ (Regime~B, singlet--triplet gap), with
95\% confidence intervals from linear regression in log-log space.
Negative $q_A$ indicates suppression; positive $q_B$ indicates
amplification, statistically significant only at $N{=}4$.
\label{tab:regimes}}
\begin{ruledtabular}
\begin{tabular}{lccc}
$N$ & dim & $q_A$ & $q_B$ \\
\colrule
4 & 70      & $-0.55 \pm 0.23$ & $+0.49 \pm 0.02$ \\
6 & 924     & $-2.49 \pm 0.35$ & $+0.02 \pm 0.12$ \\
8 & 12\,870 & $-1.28 \pm 0.41$ & $+0.02 \pm 0.15$ \\
\end{tabular}
\end{ruledtabular}
\end{table}

\section{Discussion}

The three results established here---the static no-go, the
Wilson-loop gauge obstruction, and the identification of
$R_n^{\mathrm{spin}}$ as the NOA analog---are exact and
size-independent. Together they answer three persistent questions.  \textit{Why does DFT-SOC undershoot equilibrium-route estimates?}  Because
the static polarizability vanishes by symmetry; equilibrium CISS is
intrinsically a finite-frequency effect.  \textit{Why do open-chain models give zero in
equilibrium?}  Because $\Phi_{\mathrm{SU(2)}} = 0$; a cyclic backbone
is required.  \textit{What is the correct observable?}  The spin
rotatory strength $R_n^{\mathrm{spin}}$, measurable through
magnetochiral dichroism~\cite{Rikken1997} or spin-polarized
photoabsorption.

Regime~A suppression ($q_A < 0$, Table~\ref{tab:regimes}) is robust at
all~$N$; Regime~B amplification ($q_B > 0$) is clear only at $N{=}4$
and requires larger ED or analytical treatment to resolve.

Three testable predictions follow.  (i)~Gapped chiral molecules should
show weak magnetochiral dichroism (Rikken--Raupach
protocol~\cite{Rikken1997}); diradicaloid systems are the candidates
for amplification if the Regime-B scaling survives beyond small~$N$.  (ii)~Tree-connected
$\pi$-paths should show no equilibrium CISS
[Eq.~\eqref{eq:fluxnogo}], a sharp null test.  (iii)~The enantiomer
sign flip $\sigma^{(R)}_{zz} = -\sigma^{(S)}_{zz}$
[Fig.~\ref{fig:symmetry}(a)] must hold for any equilibrium
mechanism---its violation signals transport-driven CISS.
These predictions address the impasse
identified by Naaman and Paltiel~\cite{NaamanPaltiel2026}: the spin
rotatory strength is the missing equilibrium observable their analysis
calls for---accessible via magnetochiral dichroism without transport,
explaining the chiroptical--CISS correlation through the NOA structure,
and turning each challenge they raise into a falsifiable spectroscopic
test. $R_n^{\mathrm{spin}}$ can be computed for realistic molecules
using relaxed-response
methods~\cite{HandySchaefer1984,Helgaker2000,Guerrero2025dag,Guerrero2025relax}
with SO-CASSCF/NEVPT2~\cite{Angeli2001,Neese2005}. Whether Regime-B
amplification survives in the thermodynamic limit is the central open
question; this framework makes it precisely defined, experimentally
measurable, and computationally testable.

All exact-diagonalization code and data used to produce the figures and
tables are available at [repository URL upon acceptance].

\begin{acknowledgments}
We acknowledge financial support and computational resources provided by
NeuroTechNet S.A.S.
\end{acknowledgments}



\begin{thebibliography}{22}

\bibitem{NaamanWaldeck2012}
R.~Naaman and D.~H.~Waldeck,
\textit{J. Phys. Chem. Lett.} \textbf{3}, 2178 (2012).

\bibitem{NaamanWaldeck2015}
R.~Naaman and D.~H.~Waldeck,
\textit{Annu. Rev. Phys. Chem.} \textbf{66}, 263 (2015).

\bibitem{Gohler2011}
B.~G\"ohler \textit{et~al.},
\textit{Science} \textbf{331}, 894 (2011).

\bibitem{Xie2011}
Z.~Xie \textit{et~al.},
\textit{Nano Lett.} \textbf{11}, 4652 (2011).

\bibitem{Kettner2015}
M.~Kettner \textit{et~al.},
\textit{J. Phys. Chem. Lett.} \textbf{6}, 4916 (2015).

\bibitem{Mishra2013}
D.~Mishra \textit{et~al.},
\textit{Proc. Natl. Acad. Sci. U.S.A.} \textbf{110}, 14872 (2013).

\bibitem{GuoSun2012}
A.-M.~Guo and Q.-F.~Sun,
\textit{Phys. Rev. Lett.} \textbf{108}, 218102 (2012).

\bibitem{Medina2012}
E.~Medina, F.~L\'opez, M.~A.~Ratner, and V.~Mujica,
\textit{Europhys. Lett.} \textbf{99}, 17006 (2012).

\bibitem{Matityahu2016}
S.~Matityahu, Y.~Utsumi, A.~Aharony, O.~Entin-Wohlman, and C.~A.~Balseiro,
\textit{Phys. Rev. B} \textbf{93}, 075407 (2016).

\bibitem{Michaeli2016}
K.~Michaeli and R.~Naaman,
\textit{J. Phys. Chem. C} \textbf{120}, 17402 (2016).

\bibitem{DalumHedegard2019}
S.~Dalum and P.~Hedeg\aa{}rd,
\textit{Nano Lett.} \textbf{19}, 5253 (2019).

\bibitem{Fransson2019}
J.~Fransson,
\textit{J. Phys. Chem. Lett.} \textbf{10}, 7126 (2019).

\bibitem{Fransson2020}
J.~Fransson,
\textit{Phys. Rev. B} \textbf{102}, 235416 (2020).

\bibitem{LandauLifshitz}
L.~D.~Landau and E.~M.~Lifshitz,
\textit{Electrodynamics of Continuous Media} (Pergamon, Oxford, 1984).

\bibitem{Barron2004}
L.~D.~Barron,
\textit{Molecular Light Scattering and Optical Activity}, 2nd~ed.
(Cambridge University Press, Cambridge, 2004).

\bibitem{Rikken1997}
G.~L.~J.~A.~Rikken and E.~Raupach,
\textit{Nature} \textbf{390}, 493 (1997).

\bibitem{KaneMele2005}
C.~L.~Kane and E.~J.~Mele,
\textit{Phys. Rev. Lett.} \textbf{95}, 226801 (2005).

\bibitem{Wilson1974}
K.~G.~Wilson,
\textit{Phys. Rev. D} \textbf{10}, 2445 (1974).

\bibitem{Angeli2001}
C.~Angeli, R.~Cimiraglia, S.~Evangelisti, T.~Leininger, and J.-P.~Malrieu,
\textit{J. Chem. Phys.} \textbf{114}, 10252 (2001).

\bibitem{Neese2005}
F.~Neese,
\textit{J. Chem. Phys.} \textbf{122}, 034107 (2005).

\bibitem{HandySchaefer1984}
N.~C.~Handy and H.~F.~Schaefer~III,
\textit{J. Chem. Phys.} \textbf{81}, 5031 (1984).

\bibitem{Helgaker2000}
T.~Helgaker, P.~J\o{}rgensen, and J.~Olsen,
\textit{Molecular Electronic-Structure Theory} (Wiley, New York, 2000).

\bibitem{Guerrero2025dag}
R.~D.~Guerrero,
arXiv:2607.05622 (2026).

\bibitem{Guerrero2025relax}
R.~D.~Guerrero,
arXiv:2608.06536 (2026).

\bibitem{NaamanPaltiel2026}
R.~Naaman and Y.~Paltiel,
\textit{Adv. Mater.} \textbf{38}, e2523675 (2026).

\end{thebibliography}
\end{document}